\documentclass[superscriptaddress,twocolumn,showpacs,pra,nofootinbib,longbibliography]{revtex4-1}
\usepackage{bm} 
\usepackage{quantikz}
\usepackage{graphicx} 
\usepackage{amsmath}
\usepackage{amsthm}
\usepackage{amssymb} 
\usepackage{dsfont}
\usepackage{comment}
\usepackage{amsfonts}
\usepackage{epsfig}
\usepackage{color} 
\usepackage[english]{babel}

\usepackage{mathtools}
\usepackage{multirow}
\usepackage{hyperref}
\hypersetup{
    colorlinks=true,       
    linkcolor=cyan,          
    citecolor=magenta,        
    filecolor=magenta,      
    urlcolor=cyan,           
    runcolor=cyan
}

\newcommand {\be}{\begin{equation}}
\newcommand {\ee}{\end{equation}}

\newcommand{\ba}{\begin{eqnarray}}
\newcommand{\ea}{\end{eqnarray}}

\newcommand{\ignore}[1]{}

\usepackage[margin=1in,nofoot]{geometry}

\newcommand{\beq}{\begin{equation}}
\newcommand{\eeq}{\end{equation}}
\newcommand{\beqnn}{\begin{equation*}}
\newcommand{\eeqnn}{\end{equation*}}
\newcommand{\bea}{\begin{eqnarray}}
\newcommand{\eea}{\end{eqnarray}}
\newcommand{\beann}{\begin{eqnarray*}}
\newcommand{\eeann}{\end{eqnarray*}}
\newcommand{\bes} {\begin{subequations}}
\newcommand{\ees} {\end{subequations}}

\begin{document}

\title{QuFeX: Quantum Feature Extraction Module for Hybrid Quantum-Classical Deep Neural Networks}

\author{Naman Jain}
\affiliation{Viterbi School of Engineering, University of Southern California, Los Angeles, California 90089, USA}
\author{Amir Kalev}
\email{amirk@isi.edu}
\affiliation{Information Sciences Institute, University of Southern California, Arlington, VA 22203, USA}
\affiliation{Department of Physics and Astronomy, and Center for Quantum Information Science \& Technology, University of Southern California, Los Angeles, California 90089, USA}
%
\begin{abstract}
We introduce Quantum Feature Extraction (QuFeX), a novel quantum machine learning module. The proposed module enables feature extraction in a reduced-dimensional space, significantly decreasing the number of parallel evaluations required in typical quantum convolutional neural network architectures. Its design allows seamless integration into deep classical neural networks, making it particularly suitable for hybrid quantum-classical models. As an application of QuFeX, we propose Qu-Net -- a hybrid architecture which integrates QuFeX at the bottleneck of a U-Net architecture. The latter is widely used for image segmentation tasks such as medical imaging and autonomous driving. Our numerical analysis indicates that the Qu-Net can achieve superior segmentation performance compared to a U-Net baseline. These results highlight the potential of QuFeX to enhance deep neural networks by leveraging hybrid computational paradigms, providing a path towards a robust framework for real-world applications requiring precise feature extraction.
\end{abstract}

\maketitle

\section{Introduction}

Quantum computing has emerged as a computational paradigm with the potential to solve problems intractable for classical computers. The rise of the machine learning (ML) methodology has spurred interest in quantum machine learning (QML) as a potential means to improve classification, regression, and clustering tasks~\cite{qml_survey_2020, entangled_dataset_2021}.

Early on in the development of QML, quantum neural networks (NNs) were proposed~\cite{iris_cong_qcnn,henderson_2019_quanvo} as an extension of the concept of classical NNs by utilizing quantum circuits as layers of the network. The design of the quantum layers corresponds to different ML architectures, with notable examples including Quantum Convolutional NN (QCNN)~\cite{iris_cong_qcnn} and Quanvolutional NN (QuanNN) \cite{henderson_2019_quanvo}(see also recent advancements in QuanNN~\cite{Bhatia_2023,Ceschini2025}).
While convolutional NNs (CNNs) -- the cornerstone of modern NNs --  have achieved significant success across a wide range of ML applications~\cite{cnns_survey, cnns_survey2024}, their resource requirements, especially for training large models, increase drastically with the size and complexity of data \cite{comp_limits_dl, dl_scaling_2017}.
Quantum NN extensions, such as QCNN and QuanNN, were proposed under the hypothesis that,  since quantum computers can process information in fundamentally different ways, quantum NNs may potentially offer greater efficiency compared to their classical counterparts \cite{sup_learning_with_QC, ai_in_quantum}. 

In parallel to developing quantum NN models, in recent years there has been a growing interest in creating a hybrid quantum-classical model where quantum and classical NNs work in tandem~\cite{hybrid_qc_compare, hqcnn, hqcnn_downlink, hqcnn_vqc}. This division of computing resources allows leveraging the strengths of both classical and quantum computing while managing the limitations of near-term quantum devices. Hybrid models can be particularly promising in domains such as image processing and generative modeling, where quantum circuits may offer advantages in high-dimensional feature spaces, while classical systems handle more routine tasks. Thus, the hybrid quantum-classical ML approach provides a feasible pathway toward enhancing ML tasks under the current quantum technological constraints.

\begin{figure*}[t!]
\centering
\includegraphics[width=0.85\textwidth]{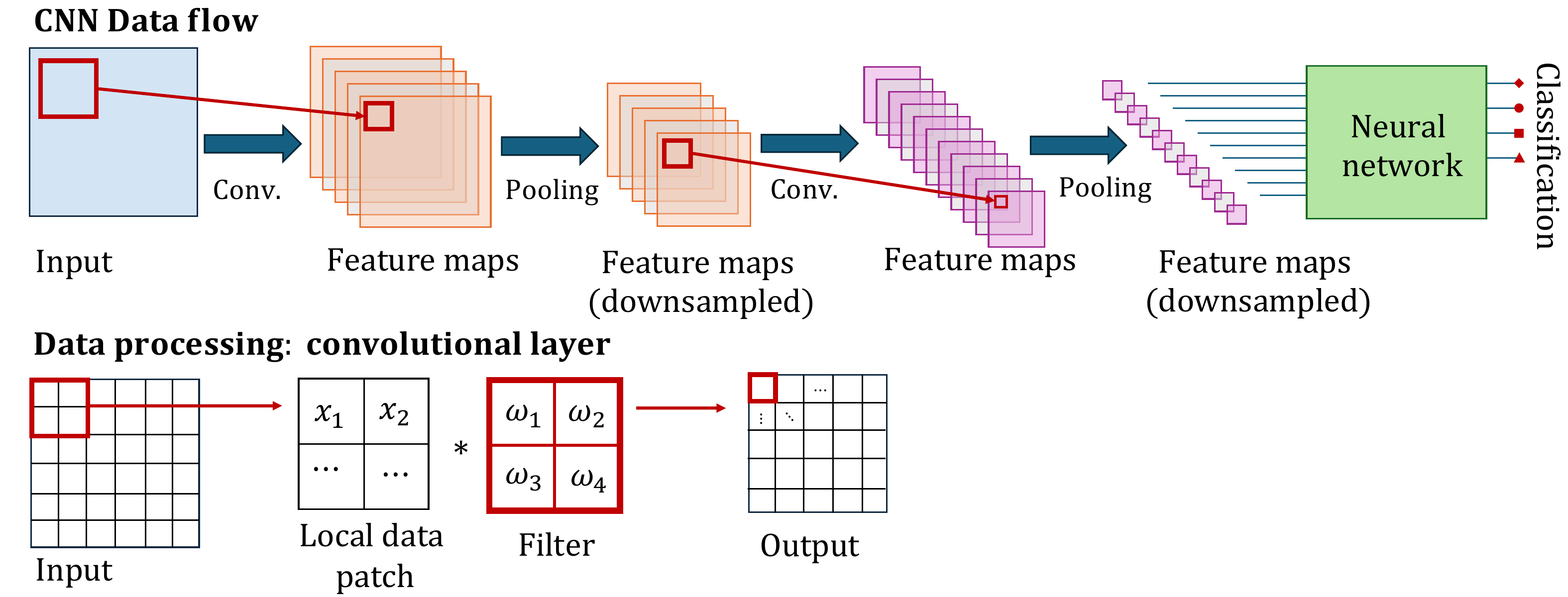}
\caption{\label{fig:convo} {\bf Typical CNN architecture}. The network consists of alternating convolutional and pooling layers for feature extraction, followed by fully connected layers usually for classification.  The figure also illustrates the convolution operation, showing how a filter interacts with the input data to produce feature maps. }
\end{figure*}

A typical hybrid model includes classical layers for pre-processing data, followed by quantum processing layers (circuits), and concludes with classical post-processing steps.  Nevertheless, various strategies have been proposed for integrating quantum layers into classical deep NN architectures. For example, in image classification tasks, quantum layers are often positioned near the end of the hybrid network to generate the final output label~\cite{Senokosov_2024}, while some approaches involve encoding data through quantum layers at the beginning of the network~\cite{img_class_hqcnn}. The design of the quantum layer itself varies widely as well, with notable examples including QCNN~\cite{iris_cong_qcnn} and QuanNN~\cite{henderson_2019_quanvo}. These variations underscore the need for a comprehensive study of optimal quantum layer placement in hybrid networks, as well as a systematic approach to constructing effective quantum layers.

In this work, we take a step forward in these two directions. We introduce a novel quantum learning architecture, which we term quantum feature extraction, or QuFeX for short, which combines and leverages techniques from QCNN~\cite{iris_cong_qcnn} and QuanNN~\cite{henderson_2019_quanvo}.  At a high-level it integrates the data analysis structure of QCNN, which was shown to excel in classification tasks~\cite{elhag2024quantumconvolutionalneuralnetworks, qnn_2018, Hur2021QuantumCN}, with the data pipeline of QuanNN which is well suited for local transformation of the input data~\cite{henderson_2019_quanvo}. This combination allows us to design a quantum layer that can extract features from the data in a reduced dimensional representation of it. This in turn results  in a reduced number of parallel quantum circuit evaluations compared to QuanNN. Thus, one of the main advantages of the QuFeX design is in its ability to be integrated in a hybrid fashion with classical layers, which makes it an excellent candidate for integration in deep classical NNs. 

As a proof of concept, we propose a novel hybrid quantum-classical model, a Qu-Net, were the QuFeX is integrated at the bottleneck of a classical U-Net model. U-Nets are a type of convolutional neural network architecture primarily designed for image segmentation tasks~\cite{unet}. They are widely used in applications such as medical imaging~\cite{unet, Yuan2024}, autonomous driving~\cite{unet_carvana}, and satellite image analysis~\cite{unet_satellite}, where precise pixel-level predictions are essential. We tested the performance of the proposed Qu-Net on three image segmentation datasets, FruitSeg30~\cite{fruitseg30}, PH\textsuperscript{2}~\cite{ph2}, and ISBI-2012 EM~\cite{isbi} and compared it to that of an all-classical U-Net baseline. Our findings, reported in detail below,  provide numerical evidence that with a careful design of quantum learning circuits, and with a strategic placement of those within deep NNs we can achieve quantum-classical hybrid learning models that leverage the strengths of both computing paradigms. Our demonstrations represent one of the first applications of quantum-enhanced deep NNs to real-world image segmentation tasks.

The remainder of the paper is organized as follows. In Section \ref{sec:QuFex}, we first present a brief overview of CNNs and their quantum counterpart architectures, setting the stage for the introduction of the QuFeX module. In Section \ref{sec:Qu-Net}, we introduce the Qu-Net architecture utilizing the  QuFeX at the bottleneck of the network, and in Sec.~\ref{sec:tests} we numerically test and analyze the performance of the Qu-Net for various model sizes and datasets and compare it to a U-Net baseline. Finally, we offer conclusions and outlook in Section \ref{sec:conclusion}.

\section{The Q\lowercase{u}F\lowercase{e}X architecture}\label{sec:QuFex}

Setting the stage for the proposed quantum architecture, we first highlight a few key components of classical CNNs, QCNNs, and QuanNNs that are relevant to our work.

\subsection{Convolutional neural networks and their quantum counterparts}
At its core, a CNN is composed of a hierarchal series of convolutional layers, which are the fundamental building blocks of modern classical NNs. As illustrated in Fig.~\ref{fig:convo} each convolutional layer computes new feature values from a weighted linear combination of nearby ones in the preceding layer. The set of weights (usually arranged in the form of a small matrix) is called a kernel or a filter. The kernel (or a collection thereof) ``slides'' across and convolve over the input data, generating a feature map. This translational invariance of the CNN makes them a powerful ML tool. The maps produced at different layers of the network highlight specific and often different patterns (features) of the input data, enabling its applicability across various learning tasks. Typically, pooling layers are interleaved between adjacent convolution layers to reduce the size of the feature maps, e.g. by taking the maximum or average value of a few contiguous feature values. A non-linear activation function, such as ReLU (rectified linear unit), usually follows the pooling layer. The weights of the kernels are optimized by training on a dataset. In contrast, variables like the number of kernels, the size of each kernel, and pooling layer specifics are called hyperparameters and are fixed for a given model. 
\begin{figure}[t!]
\centering
\includegraphics[width=0.95\columnwidth]{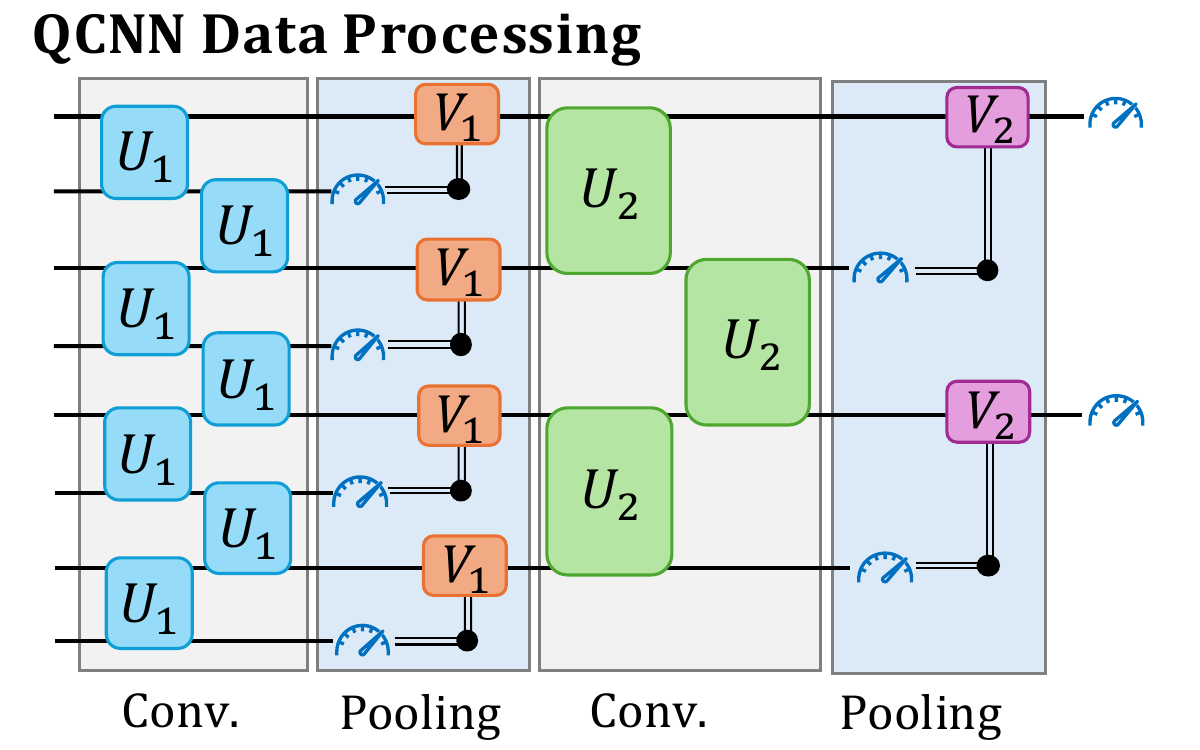}
\caption{\label{fig:qcnn} {\bf Typical QCNN data processing structure}. The QCNN integrates quantum circuits with a hierarchical structure analogous to classical CNNs, including quantum convolutional layers and pooling operations. The figure illustrates how quantum gates process input data to extract features, followed by measurement-based pooling to reduce dimensionality. }
\end{figure}

Motivated by these powerful information processing components researchers have developed quantum counterparts of CNN. QCNN, first introduced in \cite{iris_cong_qcnn}, was designed to mimic the translational invariance property of CNN. Figure~\ref{fig:qcnn} provides an illustration of the QCNN architecture. In QCNNs, the (classical) convolutional layers are replaced with a set of parametrized unitary transformations ($U_j$ in the figure) and pooling layers are replaced with fixed unitary operations ($V_j$ in the figure) conditioned on the measurement results of few specific qubits. These blocks of unitary operations (circuits) are concatenated in a serial fashion, and as shown in Fig.~\ref{fig:qcnn}, each unitary circuit is designed in a translational invariant way to maintain this CNN property. One of the advantages of this approach is that for an $N$ qubit circuit, this network has only $O(\log N)$ trainable parameters~\cite{iris_cong_qcnn}. While the structure of each circuit layer is fixed, the parametrized unitary operations ($U_j$) themselves are trainable via an optimization strategy, e.g., gradient descent to minimize a particular loss function, and therefore are often thought of as the quantum-equivalent of CNN kernels. 

While, operationally, the quantum convolutional circuits within QCNNs are considered efficient data processing modules that mimic the translational invariance of kernels in CNN, the data pipeline of QCNN generally does not possess some of the key features of CNNs. Specifically, since the convolutional and pooling layers in a typical  QCNN circuit are ordered sequentially  a circuit, the input to the circuit is often too large to be handled efficiently (this is known also as the input problem). On the other hand, classically manipulating the input to reduce it to a manageable size, may result in loss of information when handling data with multiple (or non-local) features.

\begin{figure*}[t!]
\includegraphics[width=0.85\textwidth]{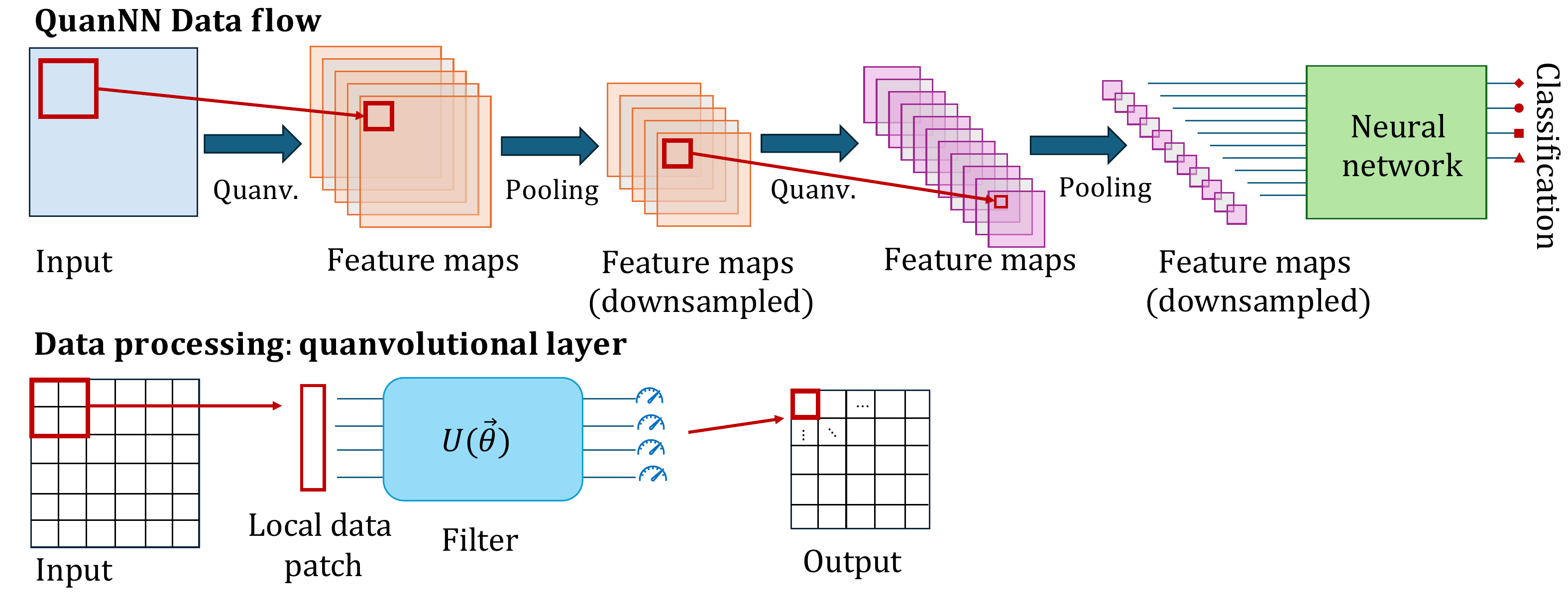}
\caption{\label{fig:quanvo} {\bf Typical QuanNN architecture}. The architecture incorporates quanvolutional layers, where classical input data is processed through quantum circuits to extract quantum-enhanced features. These features are then passed to classical layers for further processing }
\end{figure*}
To address these challenges, QuanNN~\cite{henderson_2019_quanvo} has been proposed, which leverage quantum circuits to enable the simultaneous extraction of local features across input data. QuanNNs act as feature extractors, analogous to convolutional layers in classical networks, and can be integrated seamlessly with classical NN architectures for hybrid quantum-classical processing. As illustrated in Fig.~\ref{fig:quanvo}, in QuanNN each quantum circuit is treated as a kernel and generates a feature map when applied to an input by transforming spatially-local subsections of it. These filters correspond to specific circuit designs and can be configured to potentially learn different features of the input. The QuanNN architecture thus offers greater flexibility and is considered a closer representation of classical CNNs in its way of handling the flow of data. Nonetheless, by mimicking the data flow of classical CNN,  QuanNN fall short in efficiency. Since QuanNNs extract local quantum features independently for each patch, similar to CNNs, to process the input data we must design ${\cal O}(kN)$ quantum experiments, with thousands of learning iteration per filter as reported in ~\cite{henderson_2019_quanvo}, where $N$ is the size of the input and $k$ is the number of filters used.

In this work, we take inspiration from the data processing of QCNN and the data flow in QuanNN to create a novel quantum feature tool, QuFeX, that emulate the hierarchical and local feature extraction that is characteristic to CNNs while overcoming the limitations of each architectures separately. We describe in detail next.

\subsection{QuFeX: Quantum feature extraction architecture}
\begin{figure*}[t!]
\includegraphics[width=0.9\textwidth]{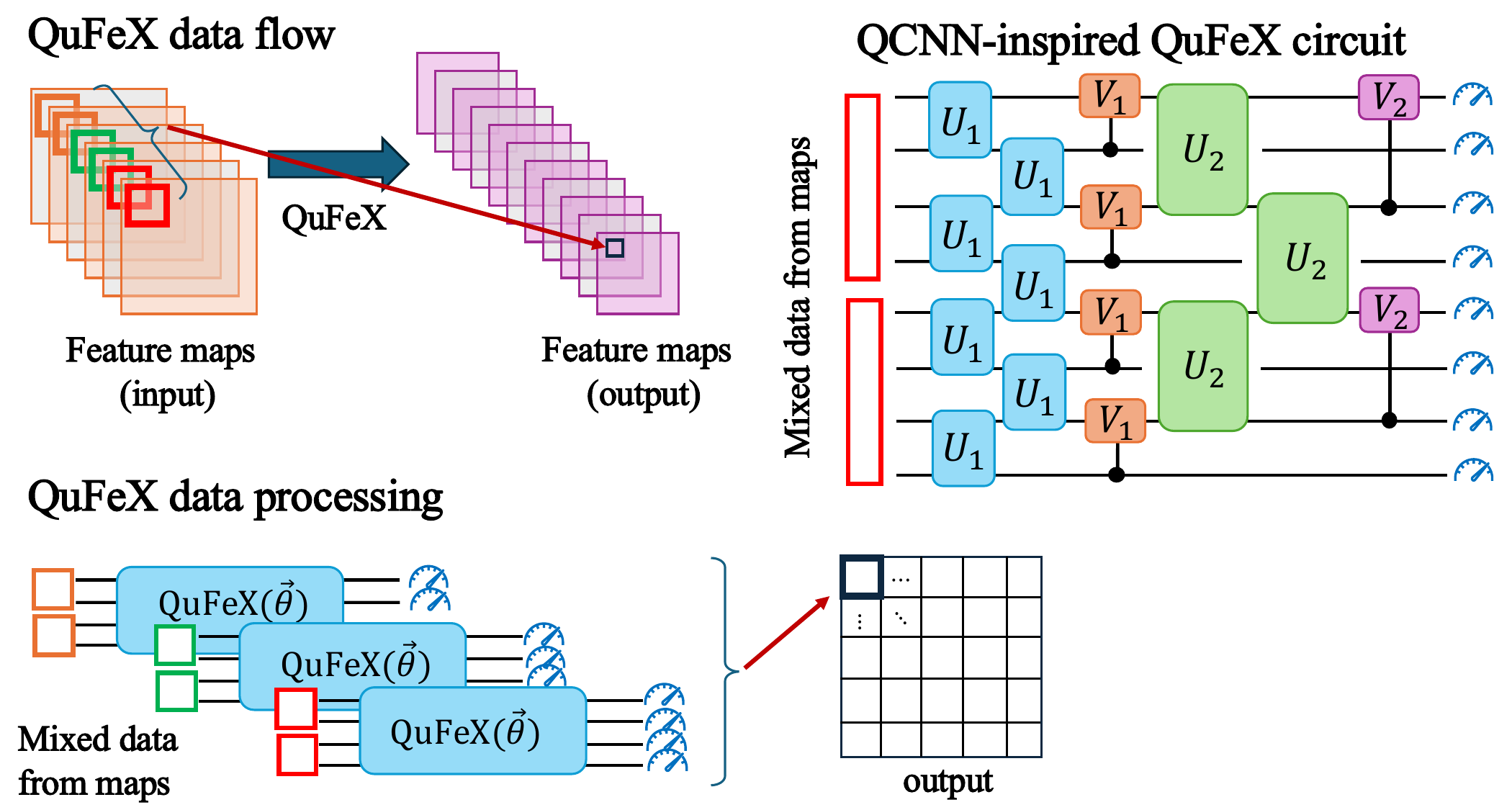}
\caption{\label{fig:QuFeX} {\bf Example of a QuFeX architecture}. The QuFeX incorporates ideas and techniques from QCNN and QuanNN. In the data flow we use information taken from different feature maps (from two maps in this example) and pass it to a convolutional-inspired quantum circuit. Example of such circuit is illustrated on the top-right part of the figure where $U_1$ and $U_2$ depend on trainable parameters (denoted in the paper by $\vec\theta$), see Appendix~\ref{app:fruit} for more details. Single-qubit measurements are then performed and the average of the corresponding Pauli-$Z$ operator is then passed on to produce an output feature map. In the QuFeX architecture we use different circuits (i.e., circuits on different qubits) to analyze different patches of the input data, in parallel. The circuits that run in parallel contain the same trainable set of parameters.}
\end{figure*}

The QuFeX architecture, shown schematically in Fig. \ref{fig:QuFeX}, follows the QuanNN style to manage the data flow and uses the QCNN spatial-invariance circuit design for data processing. In addition, we introduce several critical modifications to efficiently handle the data, specifically for near-term devices, and for compatibility with deep NNs. 

In this architecture, the QuFeX circuit is designed as a quantum filter that creates a feature map from its input data. We note by passing that while tensor-network architectures can also act as efficient classical filters through tensor decomposition, QuFeX differs conceptually and operationally: it implements a trainable quantum circuit whose filtering behavior arises from coherent operations on entangled states. Rather than claiming superiority over classical tensor-network schemes, we emphasize QuFeX as a proof-of-concept demonstration showing that shallow quantum circuits, even with few qubits, can enhance feature extraction within established deep-learning frameworks.

To avoid limiting the application of the filter to a spatially local subsection of the input data, we design it to act on a mix of different input feature maps, as illustrated in Fig.~\ref{fig:QuFeX}. This enables us to generalize the quantum feature extraction to the important (and common) cases where features are not spatially local in the input data. The way we mix the input data can be considered a hyperparameter, similar to CNN. Notably, this mixing can be done in parallel on all feature maps input, and therefore,  the QuFeX can handle $k$ input feature maps with $O(1)$ uses of the QuFeX circuit [QuFeX$(\vec{\theta})$ in the figure]. This is one of the main advantages of the proposed architecture, compared to existing ones. By using QuFeX circuit on a small number of qubits (say, 4 qubits) we are able to  handle $k$ classical feature maps, where $k$ can be very large, each of size $N\times N$, to produce an output feature map of size $M\times M$ (typically $M$ is chosen to be smaller than $N$). The cost of producing an output feature map using QuFeX depends only on the size of the QuFeX circuit and the output feature map size $M$. The exact cost may vary from one implementation to another, depending on the choice of hyperparameters such as how we mix different input feature maps and the way we slide the convolution window.   In addition,  more filters, and hence more output feature maps, can be created simply by adding multiple QuFeX layers, which can run in parallel and with each being fully customizable. 

Within each QuFeX layer we process the data in a translationally invariant way, similar to QCNNs, as depicted in the Fig.~\ref{fig:QuFeX}, with one critical change. Here, we are not  discarding the control qubits in the pooling layer. In QCNN the pooling layers are implemented by discarding the measured (control) qubits, effectively reducing the dimension of the network. In the QuFeX architecture, similar to classical CNNs and QuanNN, the reduction of dimension is done directly through the data flow. Therefore, instead of discarding the control qubits, and by doing so discarding information within a filter, we keep the control qubits and treat their output as a part of the (or even separate) feature map.  

One key advantage of the QuFeX architecture is its capacity for parallelized execution of multiple quantum filters, allowing for multi-filter operations without increasing circuit complexity. Each filter operates independently but shares a uniform execution structure. Additionally, since the data flow of QuFeX architecture closely mirrors that of CNNs, its integration with deep networks in a hybrid quantum-classical ML model is straightforward, enabling seamless compatibility with established training paradigms and architectures while leveraging quantum circuits for enhanced feature extraction.

As a proof of concept, in the next sections we propose a novel hybrid quantum-classical model, a Qu-Net, were the QuFeX is integrated within a classical U-Net model~\cite{unet}--- a CNN
with a U-shape topology specifically developed for image segmentation. We demonstrate the potential power of the QuFeX though segmentation performance of the Qu-Net architecture. Our demonstrations represent one of the first applications of quantum-enhanced deep NNs to real-world image segmentation tasks.

\section{The Q\lowercase{u}-N\lowercase{et} architecture}\label{sec:Qu-Net}
\subsection{The U-Net architecture}
Illustrated in Fig.~\ref{fig:Qu-Net}(a), the U-Net architecture is a deep  CNN with a U-shape topology specifically developed for image segmentation~\cite{unet}. It features an encoder-decoder structure with skip connections, which help preserve spatial information, and enable precise pixel-level predictions for more accurate segmentation results.  The encoder progressively downsamples the input image, capturing essential features through convolutional and pooling layers. Subsequently, the decoder upsamples the compressed feature map to the original input size, reconstructing the segmentation map. At its bottleneck, i.e., at the most compressed layers, the U-Net transforms features into a latent representation that bridges the encoder and decoder. This design ``encourages'' the model to retain only the relevant information in the bottleneck layer, facilitating the construction of an effective segmentation map.

This design helps U-Net excel in segmenting objects with fine boundaries, making it highly effective, for example, in medical applications like tumor delineation, organ segmentation, and cell tracking~\cite{unet, Yuan2024}. Beyond medical imaging, U-Net is used in various fields requiring pixel-level classification, e.g., for satellite image analysis~\cite{unet_satellite} and autonomous driving~\cite{unet_carvana}. Its ability to handle small datasets through data augmentation and its versatile performance in capturing detailed structural information has made U-Net a widely adopted tool in image segmentation across industries.

\subsection{The Hybrid Qu-Net architecture}
\begin{figure}[t!]
\includegraphics[width=0.95\columnwidth]{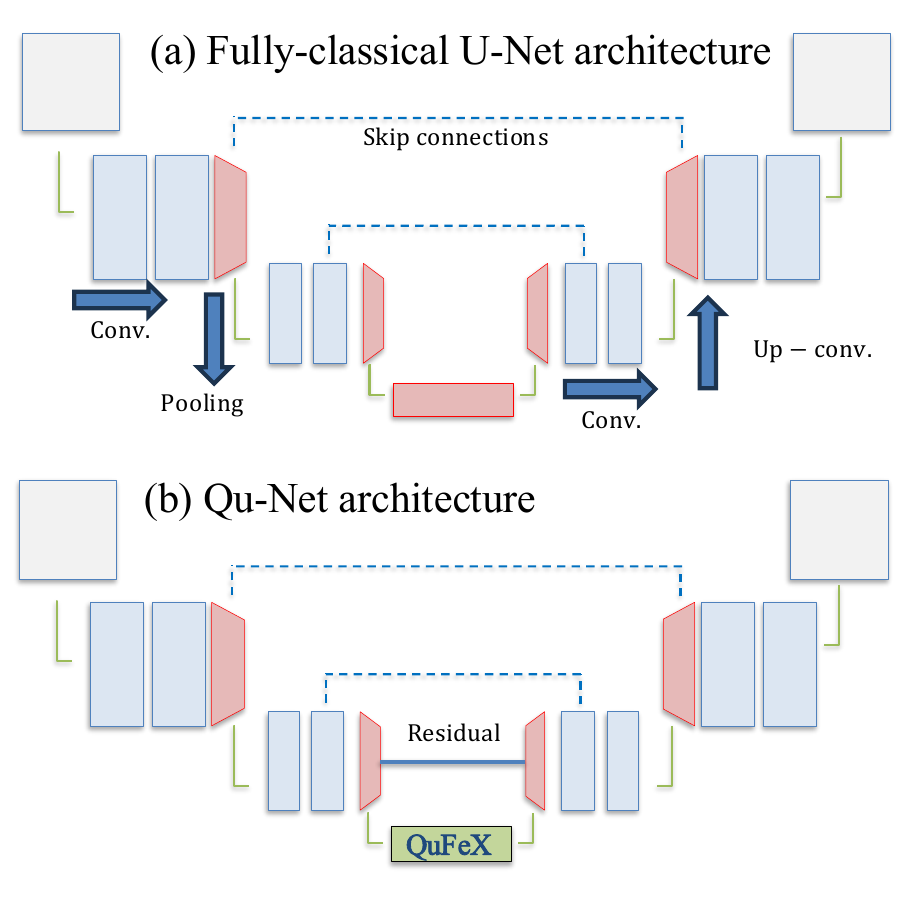}
\caption{\label{fig:Qu-Net} {\bf U-Net and  Qu-Net architectures}. (a) The U-Net architecture consists of an encoder-decoder structure with skip connections, enabling efficient feature extraction and precise localization for image segmentation. (b) The Qu-Net architecture builds on the U-Net design by replacing the classical bottleneck layer with a QuFeX module. This hybrid architecture integrates quantum circuits for enhanced feature representation while retaining the encoder-decoder structure and skip connections for segmentation tasks. }
\end{figure}
We propose a hybrid quantum-classical U-Net architecture, termed Qu-Net, that integrates the QuFeX as a quantum module within the classical pipeline, see Fig.~\ref{fig:Qu-Net}(b). Our tests and evaluation are meant to showcase the advantages of the QuFeX architecture and suggest that the hybrid model, where  QuFeX is placed at the bottleneck of the U-Net, may  deliver better image segmentation performance than compared to a baseline U-Net model. 

In hybrid models, such as the one we are considering here, the positioning of the quantum module within the classical architecture is crucial. In deep NNs, the placement of the quantum layers influences both the kinds of features that the network learns and how effectively it processes information at various stages. Different placements may impact the network’s ability to generalize, process complex correlations, and learn distinct features that might be challenging for purely classical models.

In deep CNNs early layers, closer to the input, typically capture low-level features such as edges, colors, and textures, while deeper layers identify more complex and abstract patterns. While placing a quantum layer early in the hybrid architecture might enable it to capture unique low-level properties in the raw data, potentially simplifying the task for subsequent classical layers,  this approach may under-utilize quantum resources, as the context of the entire input is limited at this stage. Similarly, positioning a quantum layer closer to the output may under-utilize potential quantum advantages particularly in cases where the classical part of the network effectively and robustly learns abstract representations of the input data. In contrast, integrating a quantum layer in the middle of the network, especially between encoding and decoding parts of deep CNNs, may present significant potential benefits. First, and most importantly, at this position the quantum layer can function similarly to a bottleneck, compressing information while preserving essential details for the final stages of processing. This strategy creates a hybrid embedding space that leverages both quantum and classical learning representations. Our tests and evaluation of the Qu-Net architecture detailed in the next subsection, support this approach. Second, from a resource perspective, placing a quantum layer in the bottleneck of a hybrid CNN, gives us the required flexibility to tailor hyperparameters, such as the size of the input to the quantum layer, to match the available quantum resources.

Hence, in the proposed hybrid U-Net we replace the bottleneck layer of the U-Net with a QuFeX.  The layers preceding and succeeding the QuFeX are (classical) U-Net layers. The level of compression (the input data to the QuFeX) is a hyperparameter chosen based on, e.g., the available quantum resources. We tested and evaluated the Qu-Net performance on image segmentation tasks and compared it to the performance of an all-classical U-Net (optimized) model.  

The  downscaling of the input all the way down to the bottleneck of the architecture,  introduces limitations on the representability of the original image. In addition to QuFeX, inspired by the ResNet architecture~\cite{resnet}, we include (classical) residual connections from one part of the network to the other that bypass the QuFeX module, see Fig.~\ref{fig:Qu-Net}(b). The residual connection can be expressed as 
$\mathbf{y} = \mathcal{Q}(\mathbf{x}) + \mathbf{x}$
where $\mathcal{Q}(\mathbf{x})$ represents the output of the quantum layer applied to the input $\mathbf{x}$ and the addition of $\mathbf{x}$ serves as the identity mapping. We argue that residual connections, typically used to stabilize training and convergence in deep NNs, are very relevant in hybrid quantum-classical setups. Adding identity mapping over the quantum layer enables the model to propagate features learned by the classical layers forward and backward through the network without being disrupted by the quantum transformation. This configuration allows the network to select an optimal path: either through a classical-only route or a path incorporating quantum processing, thus leveraging the advantages of both. Our tests below show that introducing residual connections to the Qu-Net architecture provides better segmentation performance compared to baseline U-Net models where the QuFeX is not included.

\section{Testing and evaluation}\label{sec:tests}\subsection{FruitSeg30 Image Segmentation} 
\begin{figure*}[t!]
    \centering
\includegraphics[width=0.95\textwidth]{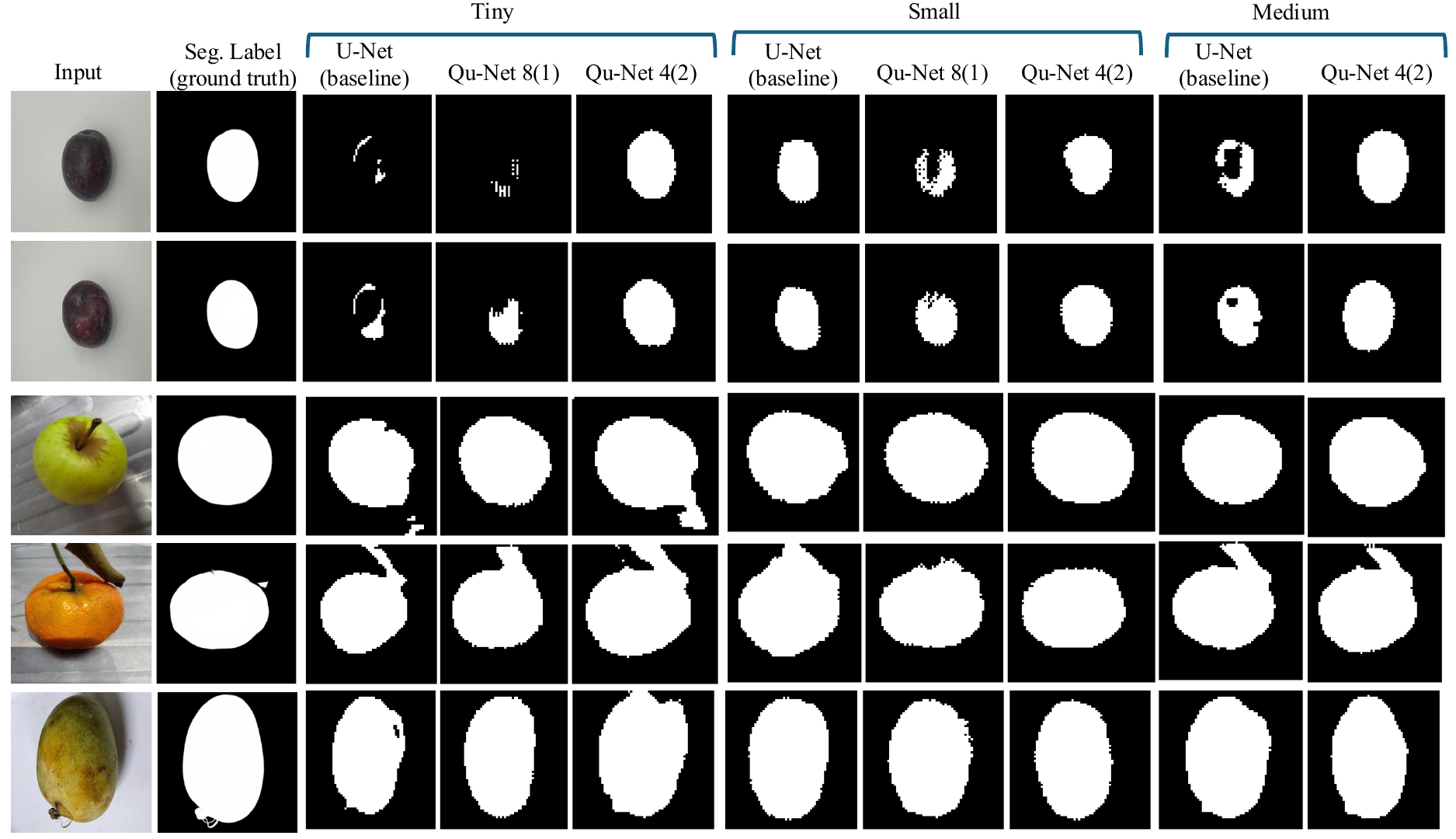}
    \caption{{\bf Qualitative comparison across models.}  The figure displays sample input images alongside their corresponding ground truth annotations and segmentation results from various model configurations. Results are shown for the baseline U-Net (fully classical) and Qu-Net variants: Qu-Net 8(1), which uses a single 8-qubit QuFeX layer, and Qu-Net 4(2), which employs two 4-qubit QuFeX layers. The comparison highlights the performance differences across model scales (tiny, small, medium) and demonstrates the impact of incorporating quantum feature extraction layers on segmentation quality.}
\label{fig:cmp-samples}
\end{figure*}

We have tested and evaluated the performance of the proposed QuFeX module within the novel Qu-Net  architecture  using the FruitSeg30 Segmentation Dataset \& Mask Annotations~\cite{fruitseg30} which consists of high-resolution images of a variety of fruits and their corresponding segmentation masks. The complete dataset includes 30 fruit classes, totaling 1969 images, each paired with precise segmentation masks. 
For this research, a curated subset of $751$ images  was selected (along with their segmentation masks) and downscaled, using Bilinear interpolation,  from the original $512 \times 512$ pixels to $64 \times 64$ pixels to align with the available computational resources [Intel Core i5-11300H CPU @ 3.10GHz (4 cores), 8 GB DDR4 RAM running Ubuntu 22.04 LTS]. The subset was selected to ensure diversity between fruit types while maintaining computational feasibility. 

To provide a baseline for the performance of the Qu-Net architecture, a U-Net architecture with approximately the same number of trainable parameters was also trained and evaluated on the same dataset. The  code for all the numerical experiments reported in this subsection are available in a public repository~\cite{Naman2025}.

We tested three variations of the models, which we refer to as tiny, small, and medium,  that consist of approximately 12,000, 26,000, and 40,000 trainable parameters, respectively. These variations in parameter count were achieved by adjusting the number of filters in the classical layers of the architectures. We refer the reader to the Appendix~\ref{app:fruit} for more details about the specific U-Net and Qu-Net models design. 

To ensure statistical reliability of the results, we generated 10 random partitions of the dataset into training and testing subsets, and run each model on these 10 pre-determined partitions. This approach accounts for potential variability in the results due to dataset splitting and ensures a robust evaluation. We use the Intersection over Union (IoU) as the performance metric for all the tested models, as it provides a balanced assessment of segmentation quality by measuring the overlap between the predicted and ground truth masks. This metric is particularly well-suited for binary segmentation tasks, ensuring accurate foreground-background separation.

Our models were trained using an Adam optimizer with a learning rate of $0.001$ and a binary cross-entropy loss function. Training was performed for 10 epochs with a batch size of 64. The Qu-Net architecture was executed on quantum simulator using PennyLane~\cite{pennylane}. An ablation study was conducted to evaluate the influence of the QuFeX layer's design choices, see the appendix~\ref{app:fruit} for more details. The performance of Qu-Net was analyzed with varying numbers of quantum filters and qubits per filter within the QuFeX module. Specifically, here we report the results obtained using one quantum filter with 8 qubits and 2 quantum filters with 4 qubits per filter. We denote those models by Qu-Net $8(1)$ and Qu-Net $4(2)$, respectively.

Figure~\ref{fig:cmp-samples} presents a qualitative assessment of the segmentation performance of the proposed Qu-Net architecture compared to the U-Net baseline. It  compares the output of Qu-Net 8(1) and Qu-Net 4(2) and the corresponding classical  U-Net baseline (alongside original image and the ground truth segmentation mask) for representative test images. These qualitative results complement the quantitative metrics, shown below, underscoring the practical advantages of Qu-Net in binary segmentation tasks. The visual improvements highlight the potential of quantum layers in enhancing feature representation and achieving superior segmentation quality.

\begin{figure}[t!]
    \centering
\includegraphics[width=0.95\linewidth]{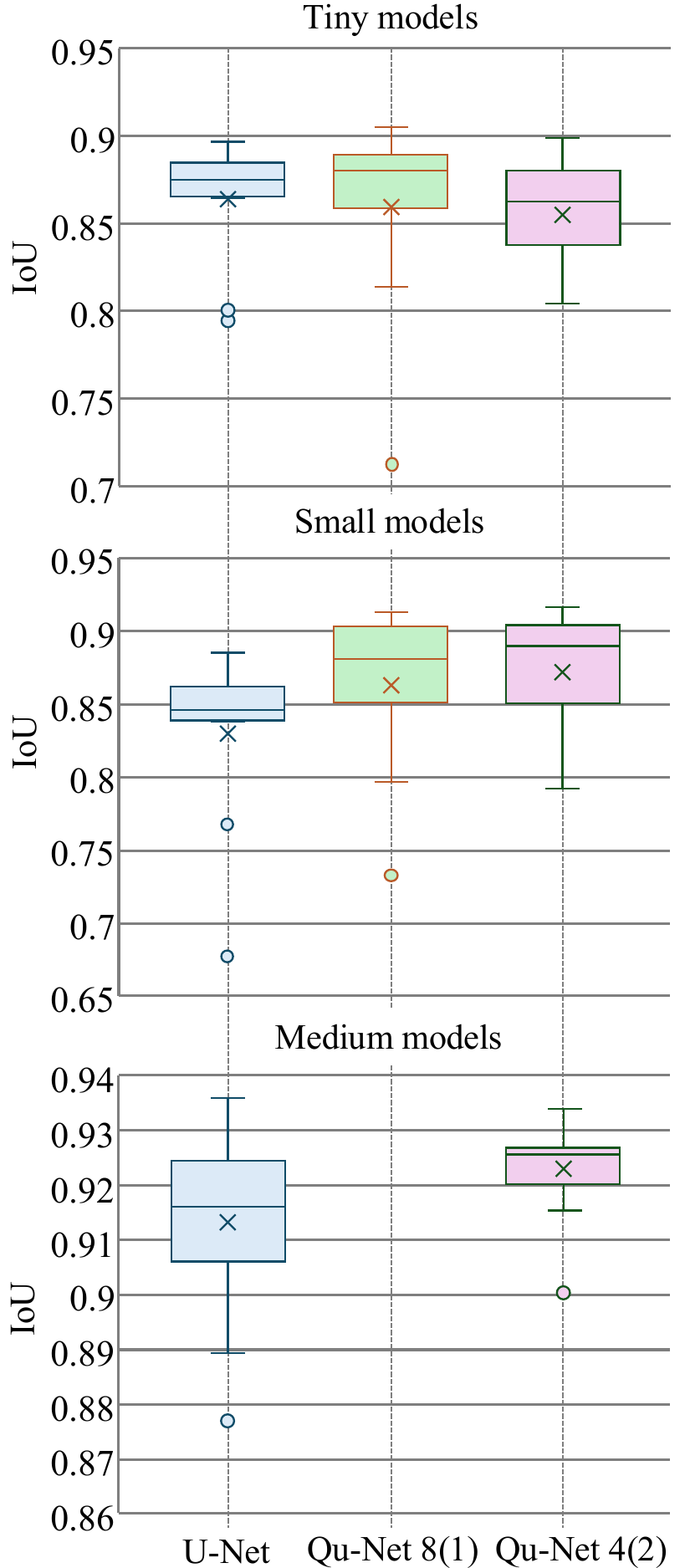}
    \caption{{\bf Comparison across models.} The data observed for different models, U-Net, Qu-Net 8(1) and Qu-Net 4(2) is summarized in a 3 boxplots for the three model variants we have considered, tiny, small and medium. Here the upper, middle and lower edges of the box represent the upper quartile ($Q_1$), median ($Q_2$) and lower  quartile ($Q_3$) of the data, respectively. The  $\times$ marks the mean IoU value, and the top (bottom) whisker extends up from the top of the box to the largest (smallest) data element that is smaller (larger) than 1.5 times the interquartile range (IQR). }
\label{fig:cmp-param}
\end{figure}
\begin{figure}[ht]
    \centering
\includegraphics[width = 0.98\linewidth]{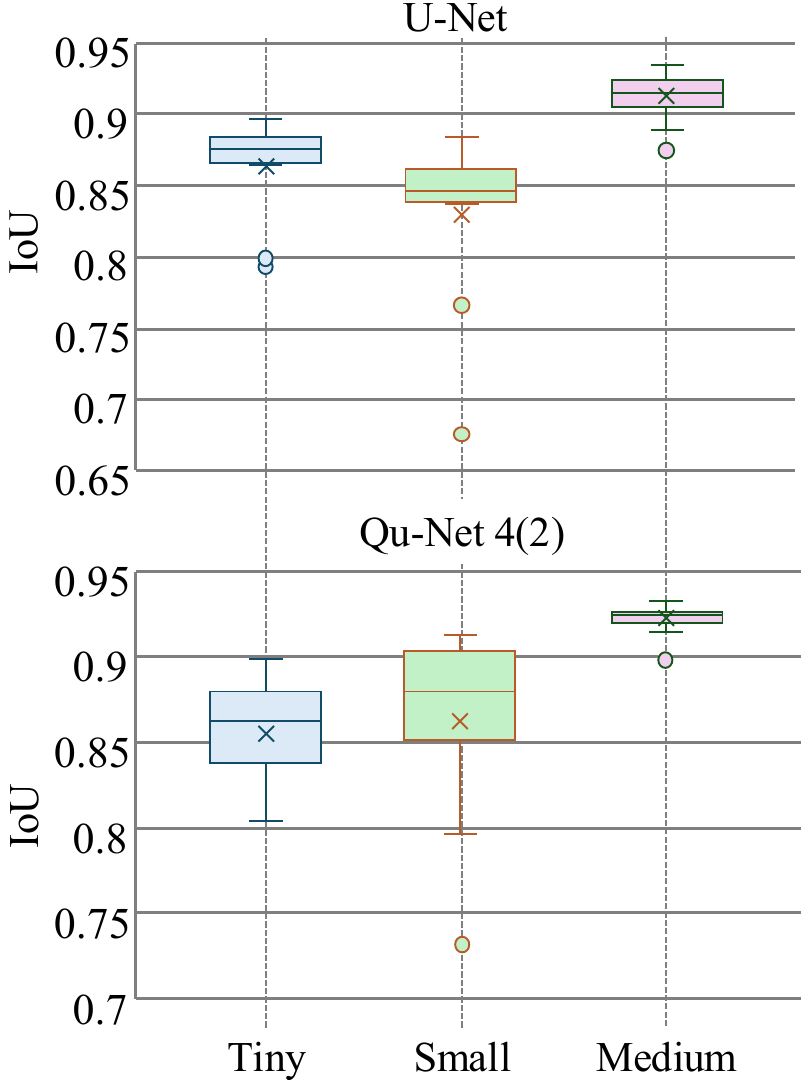}
    \caption{{\bf Comparison across model variants.} This plot summarizes the data shown in Fig.~\ref{fig:cmp-param} across model variants, tiny, small and medium. The two plots correspond to two of the three architectures we have considered, the U-Net and the Qu-Net 4(2).}
\label{fig:cmp-type}
\end{figure}
The statistical summary of these metrics is visualized in Figs.~\ref{fig:cmp-param} and~\ref{fig:cmp-type}, providing a comparative analysis of the models' performance. Each boxplot depicts the distribution of the results (IoU) from the 10 independent runs for the tested models. The median value is marked, with the interquartile range (IQR) capturing the variability of performance across different dataset partitions. Outliers are displayed to identify cases where the model's performance deviates significantly from the norm. The complete distribution of  results is given in Fig.~\ref{fig:cmp-runs} in the appendix~\ref{app:fruit}. 

Our results reveal consistent trends across the parameter regimes. While in for the tiny-type models, the U-Net outperforms the Qu-Net, in terms of median IoU and its variability across different runs, this trend is flipped as the number of tunable parameters in the models is increased.  For the small-type models the Qu-Net 8(1) and 4(2) show higher median IoU compared to their respective U-Net baseline. Nevertheless the variability of the latter is still lower than the former. However as we scale up to medium-size models, with roughly 40,000 tunable parameters, our results show that the Qu-Net provide better results than the classical baseline, in terms of both median IoU and variablilty across different runs. These results underscore the potential of integrating QuFeX with deep CNNs, and particularly with U-Net, to achieve superior segmentation performance compared to equivalent architectures that rely solely on classical methods. Figure~\ref{fig:cmp-type} further emphasizes this point by showing that while the U-Net architecture is performing qualitatively the same when scaling up the number of tunable parameters, the Qu-Net 4(2) model shows improvement  both in median IoU and IQR. We note that due to limited computational power our simulation of the Qu-Net was restricted to its 4(2) variant. 

Thus, these tests underscore  the effectiveness of Qu-Net in binary segmentation tasks, demonstrating not only superior performance but also reduced variability compared to the classical U-Net. These insights validate the robustness of the quantum-enhanced architecture to leverage quantum feature extraction for segmentation tasks.

\subsection{Medical Images Segmentation}

\subsubsection{Experimental setup}

In this section, we present a detailed account of the numerical experiments conducted on two benchmark medical image segmentation datasets to evaluate the proposed Qu-Net architecture. We compare the performance of several variants (configurations) of  Qu-Net  models against a classical U-Net baseline, establishing a benchmark for subsequent analysis. The complete implementation code and results are publicly available~\cite{Naman_Ph2,Naman_ISBI}. The quantum configurations are labeled as Qu-Net \# qubits per mixing channel-\# filters-\# channels, where the channels are  thus for example ``Qu-Net 4-1-2'' stands for a Qu-Net model variant with four qubits per mixing channel, one QuFex filtes, and two quantum mixing channels. In addition, in our experiments the segmentation task is framed as a pixel-level binary classification problem, distinguishing foreground from background pixels. To this end, we adopt the binary cross-entropy (BCE) loss function, a standard choice for such tasks \cite{Long_2015}. Moreover, the BCE was found to exhibit less overfitting compared to other candidates, such as the Dice Coefficient Loss. Our models support end-to-end training, with weights initialized randomely (see Appendix~\ref{app:medical}) and optimized using the Adam optimizer with a learning rate of $0.001$. We evaluate the models performance using the IoU metric, a gold-standard for medical segmentation tasks~\cite{ph2,isbi}. 

The several variants of the Qu-Net models correspond to modifying the number of qubits and the number of QuFeX layers, while keeping the core architecture of the QuFeX unchanged. For more details on the design of QuFeX modules, we refer the reader to the Appendix~\ref{app:medical}. We experiment with 4-, 6-, 8-, and 12-qubit configurations, both with and without quantum feature mixing. Due to limitations in computational resources required for simulating larger quantum circuits, we restrict our analysis to models with up to 12 qubits. Nevertheless, the breadth of configurations explored allows us to draw meaningful conclusions regarding performance trends and architectural efficacy.

Training the models followed a conventional supervised learning protocol. Each dataset was split into training and validation subsets. We employed batch training for a fixed number of epochs. We have not employed  extensive hyperparameter tuning since the primary objective of this work is to provide a proof of concept experiment that compare the performance of quantum-enhanced ML architecture to its classical baseline counterpart. The following subsections provide an overview of the datasets used, describe the training setup, and present results and analyses.

\subsubsection{PH\textsuperscript{2} Dataset}

The PH\textsuperscript{2} dataset~\cite{ph2} is a curated collection of 200 dermoscopic RGB images, each annotated with expert-generated segmentation masks and clinical diagnoses. The dataset includes images from three diagnostic categories: common nevus (80), atypical nevus (80), and melanoma (40). All images were resized to 192×256 pixels while preserving aspect ratio to ensure computational feasibility. To evaluate generalization and mitigate bias, we employed five-fold cross-validation with a 160/40 training/testing split. All models were trained for 20 epochs with a batch size of 18.

\begin{table}[h!] 
    \centering  
    \begin{tabular}{|l| c| c| c| c|} 
    \hline
        Model & \# Qubits & \# Param's & IoU  \\ 
        \hline\hline 
        \textbf{U-Net (baseline)} & - & \textbf{85,889} & \bm{$0.58\pm 0.09$}  \\
        Qu-Net 4-1-1 & 4 & 84,149 & $0.6 \pm 0.1$  \\
        Qu-Net 4-2-1 & 4 & 84,729 & $0.6 \pm 0.1$ \\
        Qu-Net 4-1-2 & 8 & 84,149 & $0.65 \pm 0.05$  \\
        Qu-Net 4-1-3 & 12 & 85,207 & $0.64 \pm 0.08$  \\
        Qu-Net 4-2-2 & 8 & 84,729 & $0.64 \pm 0.04$  \\
        Qu-Net 6-1-1 & 6 & 84,149 & $0.6\pm0.1$  \\
        Qu-Net 6-2-1 & 6 & 84,729 & $0.65 \pm 0.03$  \\ 
        \textbf{Qu-Net 6-1-2} & \textbf{12} & \textbf{84,149} & \bm{$0.68 \pm 0.05$}  \\
        \textbf{Qu-Net 12-1-1} & \textbf{12} & \textbf{84,149} & \bm{$0.68\pm0.04$}  \\
        Qu-Net 12-2-1 & 12 & 84,729 & $0.66 \pm 0.05$  \\
        \hline
        \hline
    \end{tabular} 
    \caption{{\bf Summary of IoU Results for PH\textsuperscript{2} Dataset.} The IoU results show the average performance over five folds and the corresponding population standard deviation rounded to the scientifically significant figures.}\label{tab:PH2_Summary}
\end{table}
Table~\ref{tab:PH2_Summary} summarizes our experimental results. It presents the models that we have tested, the number of qubits in the quantum-enhanced models, the number of training parameters per model,  a summary of the average performance, the average performance over the five folds and the corresponding population standard deviation rounded to the scientifically significant figures. The U-Net achieved an average IoU of $0.58\pm0.09$ with approximately 85,889 trainable parameters, serving as a reference for comparison. Eleven variants of the model, incorporating various architectural or algorithmic enhancements, were evaluated under the same protocol, with reported values appropriately rounded to reflect statistical precision.

Among the enhanced models, two variants, Qu-Net 6-1-2  and the Qu-Net 12-1-1, achieved the highest performance, each reporting an average IoU of $0.68 \pm 0.04$ and $0.68 \pm 0.05$, respectively. Both models exhibited a substantial increase of approximately 10\% increase over the baseline mean, while also demonstrating reduced standard deviations. This combination of higher mean accuracy and lower variance strongly suggests that these models outperform the baseline in both average and consistency of performance. Given the magnitude of improvement and the narrow confidence intervals, these enhancements are likely to be statistically significant.

Additional variants, including models Qu-Net 12-2-1, Qu-Net 6-2-1, and Qu-Net 4-2-2 reporting $0.66 \pm 0.05, 0.65 \pm 0.03$, and $0.64 \pm 0.04$, respectively, also consistently outperformed the baseline. Although the performance gains in these models are somewhat smaller, 6\% to 8\%, the lower variance suggests more robust performance across folds, indicating genuine improvements over the original U-Net.

Conversely, several models showed only marginal improvements, such as those reporting $0.6 \pm 0.1$ and $0.64 \pm 0.08$. Given the overlap in confidence intervals with the baseline, these differences are not statistically distinguishable under standard assumptions and should be interpreted conservatively. Notably, no variant performed worse than the baseline, suggesting that all modifications were either neutral or beneficial in terms of segmentation performance, while keeping the overall size of the model unchanged.

\begin{figure}[h]
\includegraphics[width=0.99\columnwidth]{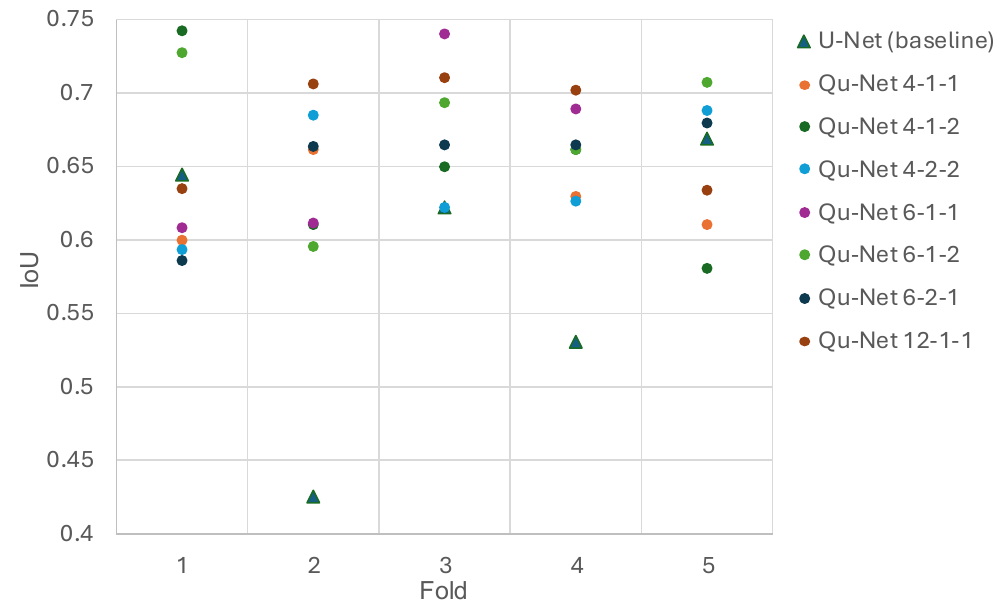}
\caption{ {\bf Representative Model Performance per Fold for the PH\textsuperscript{2} Dataset} }\label{fig:PH2-IoU-fold}
\end{figure}
In addition to mean and standard deviation metrics, we analyzed the individual per-fold IoU scores to assess the stability and reliability of each model. This per-fold view enables a nuanced evaluation of model generalization beyond aggregate statistics and helps distinguish models that achieve consistent gains from those that benefit from favorable data splits or overfit to specific partitions. A key observation is that the baseline U-Net model exhibited substantial variability across folds, with IoU scores ranging from $0.425$ to $0.669$, reinforcing its sensitivity to training data variation. Several quantum-enhanced models demonstrated improved performance not just in average IoU, but in their fold-wise stability, offering clearer evidence of robustness.  Qu-Net  6-1-2 consistently outperformed the baseline in all five folds, often by significant margins. More importantly, these models maintained tight fold-wise IoU ranges ($\leq0.04$ across four folds), suggesting that their improvements are systematic rather than stochastic. This level of consistency is particularly valuable for high-stakes applications where performance must be dependable regardless of data partitioning. Additionally Qu-Net 6-2-1 outperform the baseline in at least four folds.  By contrast, other models, including Qu-Net 4-1-1 and Qu-Net 6-1-1, showed amplified variability, with some folds significantly underperforming the baseline (e.g., IoU near or below 0.4). While their mean IoUs were comparable or slightly improved, their performance across folds was erratic, pointing to possible instability in learning dynamics or overfitting to certain data characteristics. Interestingly, a subset of models such as Qu-Net 4-2-2 and Qu-Net 4-1-2 provided modest but uniformly distributed improvements over the baseline. Though they did not achieve top performance, their consistent per-fold gains with limited variance suggest robust architecture design, possibly offering a favorable trade-off between performance and reliability.

\begin{figure}[h]
\includegraphics[width=0.99\columnwidth]{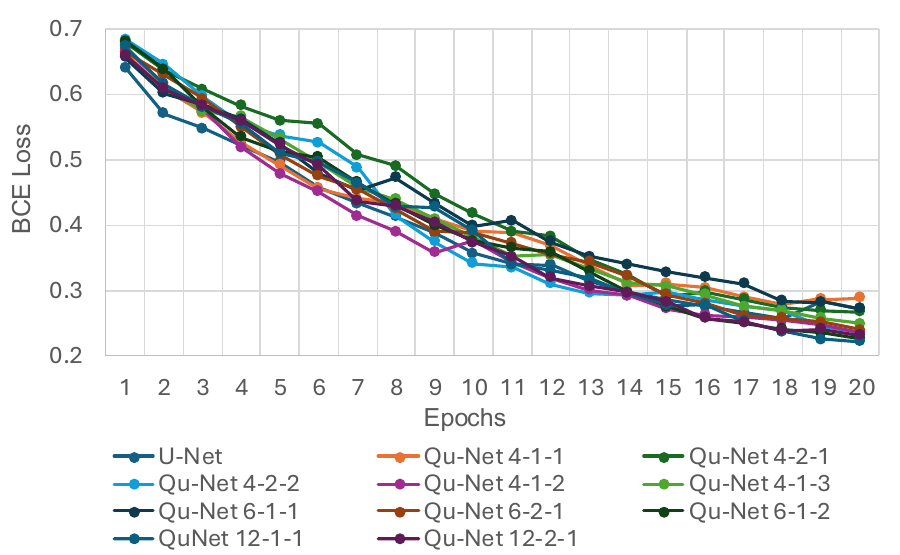}
\caption{ {\bf Average BCE Loss for the  PH\textsuperscript{2} Dataset.} }\label{fig:PH2-BCE-Av}
\end{figure}
The training dynamics, as reflected by the Binary Cross-Entropy (BCE) loss over 20 epochs, see Fig.~\ref{fig:PH2-BCE-Av}, consistently showed a decreasing trend for all trained models. These sustained reductions in BCE loss indicate that both the classical and quantum-enhanced models effectively learn to minimize the divergence between their segmentation predictions and the expert-annotated ground truths.

\begin{figure}
    \centering
    \includegraphics[width=0.99\columnwidth]{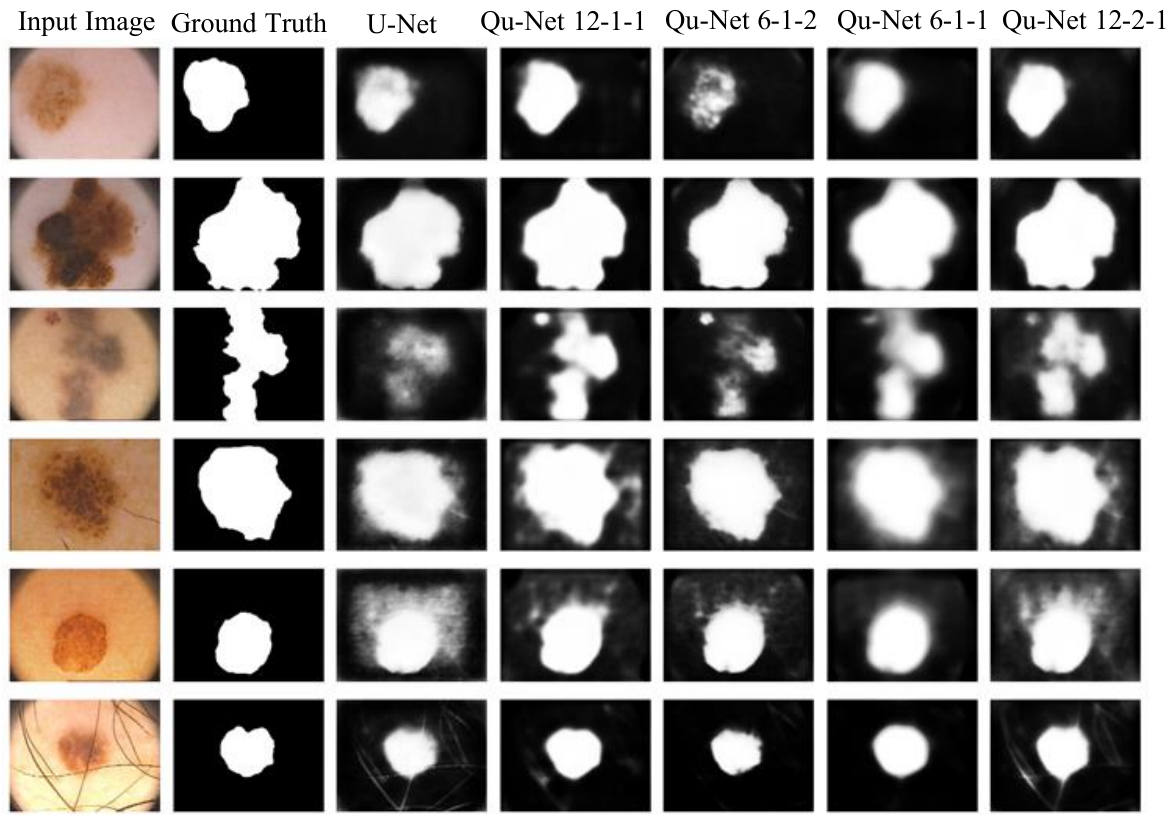}
    \caption{{\bf Sample segmentation outputs on the PH\textsuperscript{2} dataset.} From left to right: input dermoscopic image, ground truth, classical U-Net, and four Qu-Net variants. Qu-Net models, especially Qu-Net 6-1-2 and Qu-Net 12-1-1, exhibit improved boundary adherence and lesion localization compared to the classical baseline, even on atypical and faintly delineated lesions.}
    \label{fig:sample_outputs_PH2}
\end{figure}
Finally, to qualitatively evaluate the segmentation performance, in Fig.~\ref{fig:sample_outputs_PH2} we presents visual comparisons across several representative test samples. The Qu-Net models demonstrate overall better boundary preservation and lesion localization, particularly in cases involving irregular or fuzzy lesion contours. For instance, Qu-Net 6-1-2 produces segmentation masks that are visually closer to the ground truth mask, with specific instances obtaining up to $75\%$ increased accuracy, with fewer spurious predictions and sharper lesion edges than the classical U-Net. Additionally, the Qu-Net 12-1-1 variant successfully identifies melanomas with diffuse borders, highlighting the expressive advantage of quantum-enhanced latent spaces in encoding subtle diagnostic cues.

\subsubsection{ISBI-2012 EM Dataset}

The ISBI-2012 EM Segmentation Challenge dataset~\cite{isbi} comprises grayscale electron microscopy (EM) images of Drosophila melanogaster neural tissue. The task is to segment cell membranes in these high-resolution $512\times 512$ images. The dataset includes a specific fold condiguration with 30 labeled training images and 30 unlabeled test images. The primary challenges lie in the ambiguous membrane boundaries, high texture variability, and dense cellular structure. To enable training of deeper networks and expand the dataset, images were normalized and divided into $192\time 256$ pixel patches using cropping, effectively increasing the dataset size fourfold. All models were trained for 25 epochs with a batch size of 32.

\begin{table}[h!] 
    \centering 
    \begin{tabular}{|l| c| c| c| c} 
    \hline
        Model & \# Qubits & \# Param's &  IoU   \\
        \hline
        \hline
        \textbf{U-Net (baseline)}& - & \textbf{1,577,017} & \textbf{0.7545}  \\ 
        Qu-Net 4-1-1 & 4 & 259,237 & 0.7821  \\ 
        Qu-Net 4-2-1 & 4 & 259,817 & 0.7279 \\ 
        Qu-Net 4-1-2 & 8 & 259,237 & 0.7895  \\ 
        Qu-Net 4-1-3 & 12 & 261,175 & 0.8091 \\ 
        Qu-Net 4-2-2 & 8  & 259,817 & 0.7727  \\ 
        Qu-Net 6-1-1 & 6 & 259,237 & 0.7914  \\ 
        Qu-Net 6-2-1 & 6  & 259,817 & 0.8032  \\ 
        Qu-Net 6-1-2 & 12 & 259,237 & 0.7960  \\ 
        Qu-Net 12-1-1 & 12 & 259,237 & 0.7405  \\ 
        \textbf{Qu-Net 12-2-1} & \textbf{12} & \textbf{259,817} & \textbf{0.8257}\\ 
        \hline\hline
    \end{tabular} 
    \caption{{\bf Summary of  IoU Results for ISBI Dataset.} All experiments were done with 25 epochs and with a batch size of 32.} 
    \label{tab:ISBI_Summary}  
\end{table}
Table~\ref{tab:ISBI_Summary} reports the performance of various models on this dataset. The  U-Net baseline achieved a mean IoU of 0.7545, using a model with over than 1.5 million training parameters. The quantum-enhanced models demonstrated notable improvements, in terms of IoU, using an order of magnitude less of training parameters, approximately 250 thousand of them. Qu-Net 12-2-1 achieved the highest IoU of 0.8257, significantly outperforming the  baseline. Other strong-performance configurations included Qu-Net 4-1-3 (0.8091), Qu-Net 6-2-1 (0.8032), and Qu-Net 6-1-1 (0.7914). Thus these findings underpin the hypothesis that while quantum ML modules contributed a minimal number of trainable parameters they can  enhance the representational capacity of the latent space. This may be especially valuable in medical imaging, where datasets are often limited in size and rich in noise, making efficient feature extraction critical.

\begin{figure}[h]
\includegraphics[width=0.99\columnwidth]{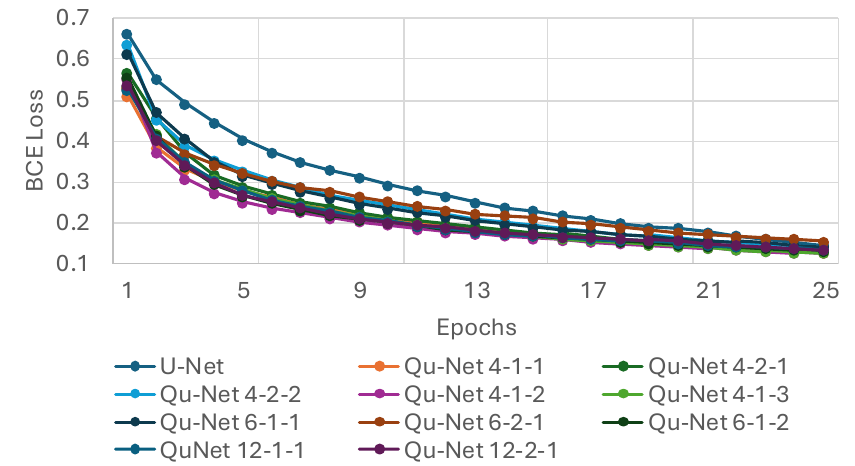}
\caption{ {\bf Average BCE Loss for the  PH\textsuperscript{2} Dataset.} }\label{fig:ISBI-BCE}
\end{figure}
To investigate optimization behavior, we monitored BCE loss throughout training for the U-Net baseline and its quantum-enhanced variants. As shown in Fig.~\ref{fig:ISBI-BCE}, all models exhibit monotonic convergence, but the learning dynamics of the baseline quantitatively differ  from those of the quantum-enhanced model. The U-Net baseline starts with the highest initial loss and maintains a consistently higher trajectory throughout all 25 epochs, indicating slower convergence and less effective optimization. In contrast, several  Qu-Net variants  achieve substantially lower BCE loss early in training and converge more rapidly to lower final losses. This suggests that architectural enhancements in these variants may improve both gradient flow and representational capacity, enabling more efficient learning. 

\begin{figure}
    \centering
    \includegraphics[width=0.99\columnwidth]{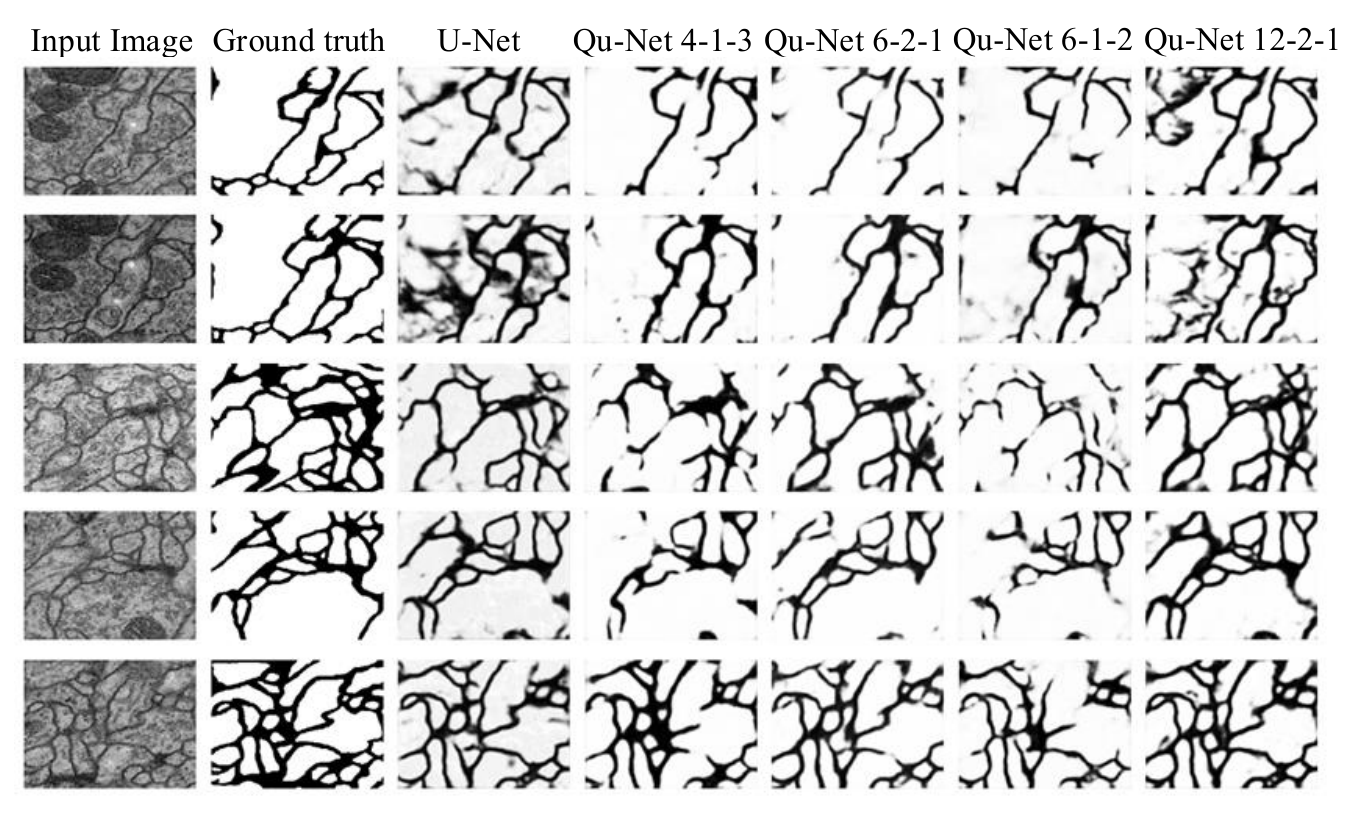}
    \caption{{ \bf Sample segmentation outputs on the ISBI-2012 segmentation dataset.} From left to right: original EM image, ground truth mask, classical U-Net, and three high-performing Qu-Net variants. Qu-Net models achieve more continuous and anatomically coherent membrane delineation, particularly in regions with ambiguous boundaries.}
    \label{fig:sample_outputs_isbi}
\end{figure}
Lastly, Fig.~\ref{fig:sample_outputs_isbi} illustrates representative outputs on the ISBI-2012 dataset. The classical U-Net often fails to capture faint or discontinuous membrane structures, resulting in fragmented boundaries or merged cell compartments. In contrast, Qu-Net variants, most notably Qu-Net 12-1-1 and Qu-Net 6-2-1, exhibit a stronger capacity to resolve intricate membrane contours. These models produce cleaner, more contiguous segmentation masks that better preserve the biological topology of neuron structures. This indicates that quantum-enhanced representations improve the network’s ability to model fine-grained spatial dependencies, which is critical in EM image segmentation where structural fidelity is paramount.

\section{Conclusion} \label{sec:conclusion}
In this work, we introduced QuFeX, a novel quantum feature extraction module designed for seamless integration into convolutional architectures. 

QuFeX is neither a quantum kernel method nor a variational classifier replacing an entire network with quantum circuits. Instead, it serves as a lightweight, trainable quantum feature-extraction layer designed to integrate seamlessly into deep convolutional pipelines. This makes QuFeX a practically viable and theoretically distinct contribution to hybrid quantum-classical learning, enabling end-to-end optimization while minimizing circuit evaluations.

By drawing inspiration from QCNN~\cite{iris_cong_qcnn} and QuanNN~\cite{henderson_2019_quanvo}, QuFeX combines efficient data flow with expressive quantum representations, positioning it as a lightweight yet powerful module for hybrid quantum-classical deep learning. A central advantage of QuFeX lies in its efficiency and scalability compared to existing approaches. QCNN architectures, while effective in introducing translational invariance, typically require qubit pooling operations that progressively reduce the representational capacity of the network, leading to information loss. In contrast, QuFeX avoids qubit discarding altogether, preserving richer intermediate features while maintaining shallow circuit depth. Similarly, QuanNN-based designs rely on repeated patchwise evaluations, with computational cost scaling as $O(kN)$ for $k$ filters and $N$ patches. QuFeX, by contrast, mixes input feature maps in a parallelizable fashion before quantum embedding, allowing the creation of multiple quantum filters with only a constant number of circuit executions. Crucially, the number of trainable parameters in QuFeX is independent of the number of qubits, decoupling model capacity from hardware limitations and ensuring favorable scaling to larger input sizes and deeper hybrid architectures. These properties make QuFeX uniquely suited for deployment on near-term devices with limited qubit counts, while also laying the foundation for scalability as quantum hardware matures. 

We note that while tensor-network architectures, such as~\cite{Stoudenmire2016,Sun2020}, can also act as efficient classical filters through tensor decomposition, QuFeX differs conceptually and operationally: it implements a trainable quantum circuit whose filtering behavior arises from coherent operations on entangled states. Rather than claiming superiority over classical tensor-network schemes, we emphasize QuFeX as a proof-of-concept demonstration showing that shallow quantum circuits, even with few qubits, can enhance feature extraction within established deep-learning frameworks.

We evaluated QuFeX by integrating it into the bottleneck of a U-Net, yielding a family of hybrid models collectively referred to as Qu-Net. Across diverse segmentation tasks, from fruit images to clinically relevant benchmarks such as PH\textsuperscript{2} dermoscopy and ISBI-2012 electron microscopy, Qu-Net consistently outperformed the classical U-Net baseline. Importantly, these gains were achieved with significantly fewer trainable parameters, underscoring the efficiency of quantum-enhanced feature extraction. On PH\textsuperscript{2}, Qu-Net variants improved mean IoU by up to 10\% with enhanced boundary preservation, while on ISBI-2012 Qu-Net 12-2-1 achieved an IoU of 0.8257 using only $\sim$250k parameters, compared to 0.7545 for U-Net with over 1.5M parameters.

While our numerical tests  use ideal quantum operations, current quantum computing devices are subject to various forms of  noise that can significantly degrade quantum circuit performance and negatively affect the end result. The impact of noise on the QuFeX module's ability to extract robust features is an active area for investigation. To ensure QuFeX's viability on current hardware, our ongoing work explores noise-mitigation techniques, including noise-aware training where we incorporate noise models during the training phase to make the QuFeX parameters more robust to hardware imperfections; applying post-training error mitigation techniques such as Zero-Noise Extrapolation (ZNE)~\cite{Temme2017,Giurgica2020} and Probabilistic Error Cancellation (PEC)~\cite{Temme2017,Zhang2020,Sun2021}; and integrating mid-circuit gate operations, such as Dynamical Decoupling (DD) gate sequences~\cite{Pokharel18}, that can result in quantum circuits that are inherently more resilient to specific noise channels prevalent in current quantum processors. Looking forward, implementing Qu-Net on near-term quantum hardware and further optimizing quantum circuits for hardware-aware deployment present exciting avenues for future research. In addition to segmentation tasks be plan to explore the applicability of QuFeX for quantum information tasks, such as entanglement detection~\cite{Chen_2022,Zhang2023}.

\section*{ACKNOWLEDGEMENTS}
This work was supported in part by an AI4Health Award from the University of Southern California and by a gift from the Alan J. Arnold, MD Foundation.

\appendix
\section{Details on the numerical tests: FruitSeg30}\label{app:fruit}
The U-Net architecture (see Fig. \ref{fig:Qu-Net}a) used in this study consists of an encoder-decoder structure with symmetric skip connections to preserve spatial information during upsampling. In our numerical tests, the input to the U-Net is 2D images of size $64 \times 64$ and its output is segmentation masks of the same dimension. The encoder comprises five convolutional blocks. Each block includes two convolutional layers with $3 \times 3$ filters, and ReLU activation. Max-pooling with a $2\times 2$ window and stride of 2 is applied after each block for downsampling. The decoder mirrors the encoder, consisting of five upsampling blocks. Each block includes an upsampling operation (transposed convolution with $2\times2$ filters), concatenation with the corresponding encoder feature map, followed by two convolutional layers with $3\times3$ filters, and ReLU activation. Skip connections are implemented by concatenating feature maps from the encoder to the corresponding decoder layers at the same resolution, ensuring preservation of spatial context. The final layer applies a  $1\times1$ convolution with a single filter and a sigmoid activation function to produce the segmentation mask. 

Unlike standard U-Net implementations, the number of filters in the convolutional blocks does not double after each block. Instead, the filter configuration is tailored to balance model complexity and performance. In the tiny scale, the encoder consists of five convolutional blocks with the number of filters set to 4, 4, 8, 8, and 8, followed by 4 filters in the bottleneck layer. In the small scale, the encoder uses 4, 8, 8, 8, and 16 filters, with 8 filters in the bottleneck. For the medium scale, the encoder is configured with 8, 8, 8, 16, and 16 filters, and the bottleneck layer contains 16 filters. In all cases, the decoder mirrors the encoder, ensuring symmetry in feature extraction and reconstruction. This design provides a flexible framework to adapt the model's parameter count to different resource constraints and application needs.

The Qu-Net design follows a structure similar to the U-Net, with identical encoder and decoder layers. The distinction lies in the bottleneck layer, where the classical bottleneck is replaced by a QuFeX layer. In the simplest configuration with one QuFeX layer, an 8-qubit quantum circuit is employed. For configurations with two QuFeX layers, the model uses 4-qubit circuits for each layer. The process of data transformation from classical to quantum, the parameterized quantum circuits, and the measurement strategy are detailed later in the text.

For instance, in the tiny scale with one QuFeX layer, the encoder outputs 8 feature maps, each of size $2\times2$. These feature maps are grouped into 4 groups, with 2 feature maps per group. Each group is then processed by an 8-qubit QuFeX layer. For configurations with two QuFeX layers [Qu-Net 4(2)], no grouping is performed (i.e., in this case there is no feature map mixing); all 8 feature maps of size $2\times2$ are fed directly to both QuFeX layers. The same treatment is extended to the small and medium scales. The total number of parameters in all the model types are listed in Table~\ref{tab:num-params}.

\begin{table}[]
\begin{tabular}{|c|c|c|c|}
\hline
 & \textbf{QU-Net} & \textbf{Qu-Net 8(1)} & \textbf{Qu-Net 4(2)} \\ \hline
\textbf{Tiny} & 12085 & 12081 + 4 & 12657 + 8 \\ \hline
\textbf{Small} & 24533 & 24525 + 4 & 26829 + 8 \\ \hline
\textbf{Medium} & 39689 & n/a & 39673 + 8 \\ \hline
\end{tabular}
\caption{{\bf Number of trainable parameters per model}. For the hybrid models we use the convention $X+Y$ where $X$ is the number of parameters in the classical part of the model and $Y$ is the number of parameters in the quantum part of it. }
\label{tab:num-params}
\end{table}

It is important to emphasize that these configurations are fully customizable, and the choices presented here are driven by constraints on computational resources. This design provides flexibility in adapting the quantum layers to different scales and hardware limitations.

\begin{figure}[t!]
    \centering
\begin{quantikz}
        \lstick{} & \gate[2]{U_1} &   \\
        \lstick{} &   & 
\end{quantikz}=\begin{quantikz}
        \lstick{} & \gate{RX(\theta_1)} & \ctrl{1} &  \\
        \lstick{} & \gate{RZ(\theta_2)} & \targ{} & 
    \end{quantikz}
\begin{quantikz}
        \lstick{} & \gate[2]{U_2} &   \\
        \lstick{} &   & 
\end{quantikz}=\begin{quantikz}
        \lstick{} & \gate{RX(\theta_3)} & \ctrl{1} &  \\
        \lstick{} & \gate{RY(\theta_4)} & \targ{} & 
    \end{quantikz}
     \caption{{\bf $U_1$ and $U_2$ 
 of a single QuFeX layer used in our numerical analysis.}}
    \label{fig:Us}
\end{figure}
The QuFeX layer is a parameterized quantum layer designed to replace the classical bottleneck layer in the Qu-Net architecture. For a single QuFeX layer using an 8-qubit circuit, there are 4 trainable parameters. The classical inputs are scaled by $\pi$ and encoded as rotation angles for 
$RY$ gates (angle encoding). The parameterized gates in the circuit are represented in Fig. \ref{fig:Us}. The pooling gates ($V_j$'s) are control-$Z$ gates in all implementation. The outputs are obtained as the expectation values of the Pauli-$Z$ observable on each qubit.

In the configuration with two QuFeX layers, the design for the second layer differs slightly. It also contains 4 trainable parameters, but the input data is encoded using $RZ$ gates with angle encoding in the $X$ basis. Additionally, the parameterized gates remain the same except that $U_1$ and $U_2$ blocks are interchanged. The measurement strategy remains consistent, extracting the expectation values of the Pauli-$Z$ observable.

It is important to note that these constructions are merely design choices and do not impose any inherent constraints. The configurations are fully modifiable and can significantly influence the model's performance. 
\begin{figure}[h!]
   \centering
\includegraphics[width=0.98\linewidth]{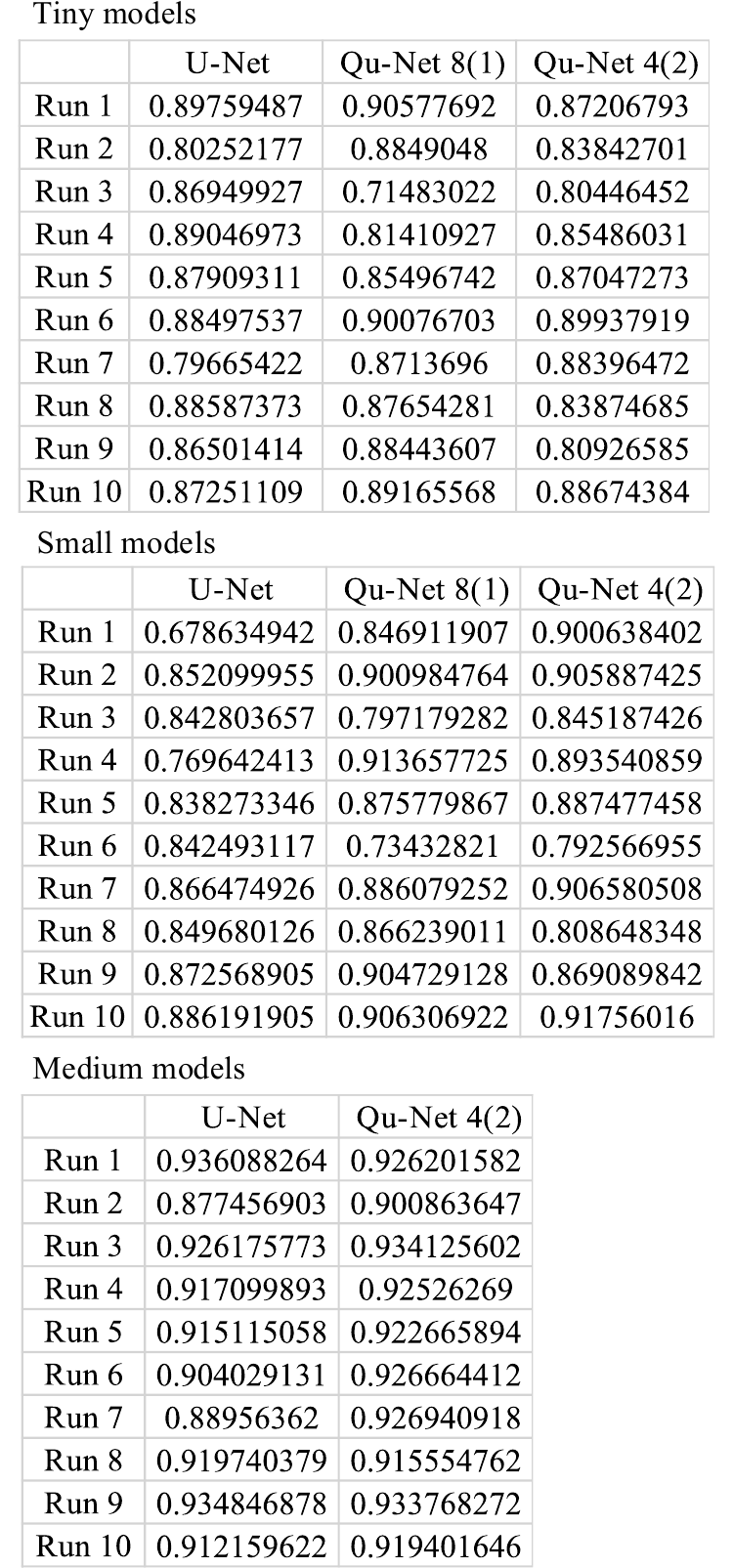}
    \caption{{\bf Raw data from numerical experiments (IoU).} }
    \label{fig:cmp-runs}
\end{figure}
In Fig.~\ref{fig:cmp-runs} we provide the run-wise IoU scores (raw data) for all performed tests.

\section{Details on the numerical tests: PH\textsuperscript{2} and ISBI-2012 EM}\label{app:medical}
\begin{figure}[h!]
    \centering
\begin{quantikz}
        \lstick{} & \gate[2]{U_1} &   \\
        \lstick{} &   & 
\end{quantikz}=\begin{quantikz}
        \lstick{} & \gate{RX(\theta_1)} & \ctrl{1} &  \\
        \lstick{} & \gate{RZ(\theta_2)} & \targ{} & 
    \end{quantikz}
\begin{quantikz}
        \lstick{} & \gate[2]{U_2} &   \\
        \lstick{} &   & 
\end{quantikz}=\begin{quantikz}
        \lstick{} & \gate{RX(\theta_3)} & \ctrl{1} &  \\
        \lstick{} & \gate{RY(\theta_4)} & \targ{} & 
    \end{quantikz}
     \caption{{\bf $U_1$ and $U_2$ 
 of a single QuFeX layer used in our numerical analysis.}}
    \label{fig:Us}
\end{figure}
The U-Net architecture (see Fig. \ref{fig:Qu-Net} (a)) used in this study consists of an encoder-decoder structure with symmetric skip connections to preserve spatial information during upsampling. In our numerical tests, the input to the U-Net is a $192 \times 256$ image, and its output is a segmentation mask of the same dimension. The encoder comprises five convolutional blocks, each containing two convolutional layers with $3 \times 3$ filters and ReLU activation. Max-pooling with a $2 \times 2$ window and a stride of 2 is applied after each block for downsampling. The decoder mirrors the encoder and consists of five upsampling blocks. Each block includes a transposed convolution with $2 \times 2$ filters for upsampling, followed by concatenation with the corresponding encoder feature map, and two $3 \times 3$ convolutional layers with ReLU activation. Skip connections are implemented via concatenation of feature maps from the encoder to the decoder at matching resolutions, ensuring the preservation of spatial context. The final layer applies a $1 \times 1$ convolution with a single filter and a sigmoid activation function to produce the segmentation mask.

For the PH$^2$ dataset, the number of convolutional filters in the encoder layers is configured as $\{8, 16, 32, 16, 8\}$, with a corresponding mirrored structure in the decoder. In contrast, for the ISBI dataset, Batch Normalization layers are added after each convolutional layer to improve training stability and generalization, and the number of convolutional filters in the encoder follows the progression $\{8, 16, 32, 64, 128\}$. This modification results in a deeper bottleneck and richer feature representation for the ISBI case, while the PH$^2$ configuration is designed to be more lightweight and dataset-specific.

The Qu-Net design follows a structure similar to the U-Net, with identical encoder and decoder layers. The distinction lies in the bottleneck layer, where the classical bottleneck is replaced by a QuFeX layer. In the simplest configuration with one QuFeX layer, a 4-qubit quantum circuit is employed. For configurations with two QuFeX layers, the model uses 4-qubit circuits for each layer. 

The QuFeX layer is a parameterized quantum layer designed to replace the classical bottleneck layer in the Qu-Net architecture. For a single QuFeX layer using an 8-qubit circuit, there are 4 trainable parameters. The classical inputs are scaled by $\pi$ and encoded as rotation angles for 
$RY$ gates (angle encoding). The parameterized gates in the circuit are represented in Fig. \ref{fig:Us}. The pooling gates ($V_j$'s) are control-$Z$ gates in all implementations. The outputs are obtained as the expectation values of the Pauli-$Z$ observable on each qubit.

In the configuration with two QuFeX layers, the design for the second layer differs slightly. It also contains 4 trainable parameters, but the input data is encoded using $RZ$ gates with angle encoding in the $X$ basis. Additionally, the parameterized gates remain the same except that $U_1$ and $U_2$ blocks are interchanged. The measurement strategy remains consistent, extracting the expectation values of the Pauli-$Z$ observable.

It is important to note that these constructions are merely design choices and do not impose any inherent constraints. The configurations are fully modifiable and can significantly influence the model's performance. 

\bibliography{refs}
\end{document}